\documentclass[journal]{IEEEtran}
\usepackage{changes}

\usepackage{amsmath,epsfig}
\usepackage{pbox}
\usepackage{tikz}
\usetikzlibrary{arrows.meta,calc,decorations.markings,math,arrows.meta,positioning}
\usepackage{pgfgantt}
\usepackage{gensymb}
\usepackage{amsmath}
\usepackage{adjustbox} 
\usepackage{amsfonts}

\usepackage{hyperref}
\usepackage{multirow}
\usepackage{todonotes}

\usepackage{array}
\usepackage{graphicx}
\usepackage{multirow}

\usepackage{pgfplots}

\usepackage{algorithm}
\usepackage{algpseudocode}

\usepackage{soul}

\ifCLASSINFOpdf
\else
\fi
\usepackage{pbox}
\usepackage{tikz}
\usepackage{pgfgantt}
\usepackage{subfigure}
\usepackage{gensymb}
\usepackage{amsmath}

\usepackage{array}
\usepackage{graphicx}
\usepackage{multirow}
\usepackage{xcolor,colortbl}
\usepackage[normalem]{ulem}

\usetikzlibrary{matrix,shapes,arrows,positioning,chains}
\tikzstyle{process} = [rectangle, minimum width=3cm, minimum height=1cm, text centered, draw=black]
\tikzstyle{arrow} = [thick,->,>=stealth]
\tikzstyle{io} = [trapezium, trapezium left angle=70, trapezium right angle=110, text centered]

\makeatletter
\def\@copyrightspace{\relax}
\makeatother

\begin{document}
%
\title{Confidence Estimation in Unsupervised Deep Change Vector Analysis}
%
%
%

\author{
        Sudipan~Saha
\thanks{}
\thanks{Sudipan Saha is with Yardi School of Artificial Intelligence, IIT Delhi, New Delhi,  India. (E-mail: sudipan.saha@scai.iitd.ac.in) }
}

\markboth{}%
{Shell \MakeLowercase{\textit{et al.}}: Bare Demo of IEEEtran.cls for IEEE Journals}
%

\maketitle

\begin{abstract}
Unsupervised transfer learning-based change detection methods exploit the feature extraction capability of pre-trained networks to distinguish changed pixels from the unchanged ones. However, their performance may vary significantly depending on several geographical and model-related aspects. In many applications, it is of utmost importance to provide trustworthy or confident results, even if over a subset of pixels. The core challenge in this problem is to identify changed pixels and confident pixels in an unsupervised manner. To address this, we propose a two-network model - one tasked with mere change detection and the other with confidence estimation. While the change detection network can be used in conjunction with popular transfer learning-based change detection methods such as Deep Change Vector Analysis, the confidence estimation network operates similarly to a randomized smoothing model. By ingesting ensembles of inputs perturbed by noise, it creates a distribution over the output and assigns confidence to each pixel's outcome. We tested the proposed method on three different Earth observation sensors: optical, Synthetic Aperture Radar, and hyperspectral sensors. 
\end{abstract}


\fboxsep=0mm
\fboxrule=0.1pt

%
\IEEEpeerreviewmaketitle

\bstctlcite{IEEEexample:BSTcontrol}

\section{Introduction}
\label{sectionIntroduction}

Change detection (CD) in bi-temporal images is an important application in Earth observation. As our planet is evolving rapidly, change detection plays an important role in many downstream applications, e.g., urban monitoring \cite{saha2019unsupervised,song2022mstdsnet}, environmental monitoring \cite{cardille2022multi,khankeshizadeh2022fcd}, and analyzing the effects of natural disasters \cite{qing2022operational,hamidi2023fast}. Advancements in machine/deep learning have propelled a diverse range of enhancements in artificial intelligence for Earth observation, including for change detection. However, deep learning based supervised change detection methods are often hindered by the lack of annotated training data. This gap is filled by deep transfer learning based methods, e.g., Deep Change Vector Analysis (DCVA) \cite{saha2019unsupervised}, that employs a pre-trained network as bi-temporal zero-shot feature extractor. As shown in \cite{saha2021trusting}, deep transfer learning based methods provide more trustworthy result than the supervised ones when training data is not abundant.
\par

\tikzset{
    between/.style args={#1 and #2}{
         at = ($(#1)!0.5!(#2)$)
    }
}

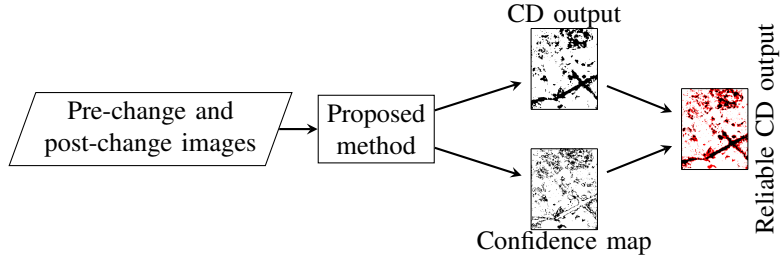
\begin{figure*}
 \centering
 \begin{tikzpicture}
\node (1) [draw, io, align=center] {Pre-change and\\ post-change images};

\node (2) [draw, right of=1, xshift=2cm, align=center] {Proposed\\ method};

\node (3a) [right of=2, xshift=1.5cm, yshift=+0.8cm] {\fbox{\includegraphics[width=25pt]{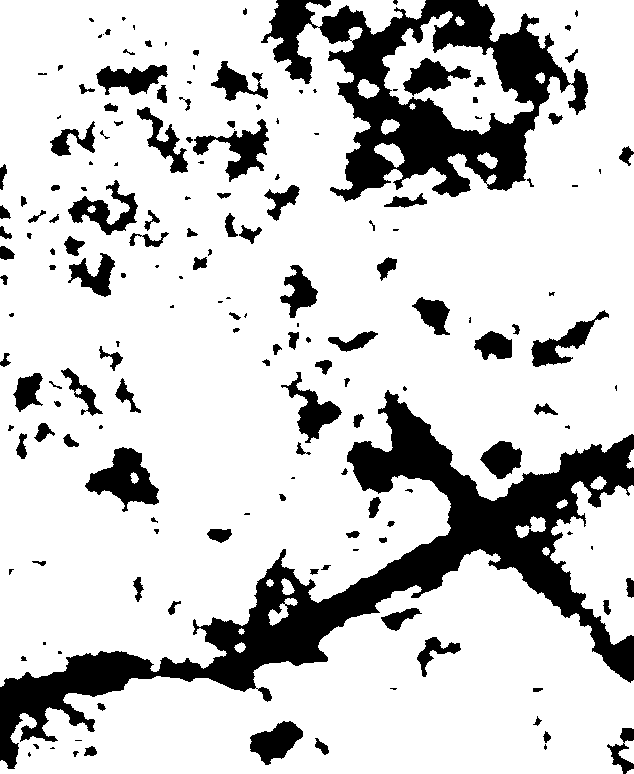}}};

\node (3b) [right of=2, xshift=1.5cm, yshift=-0.8cm] {\fbox{\includegraphics[width=25pt]{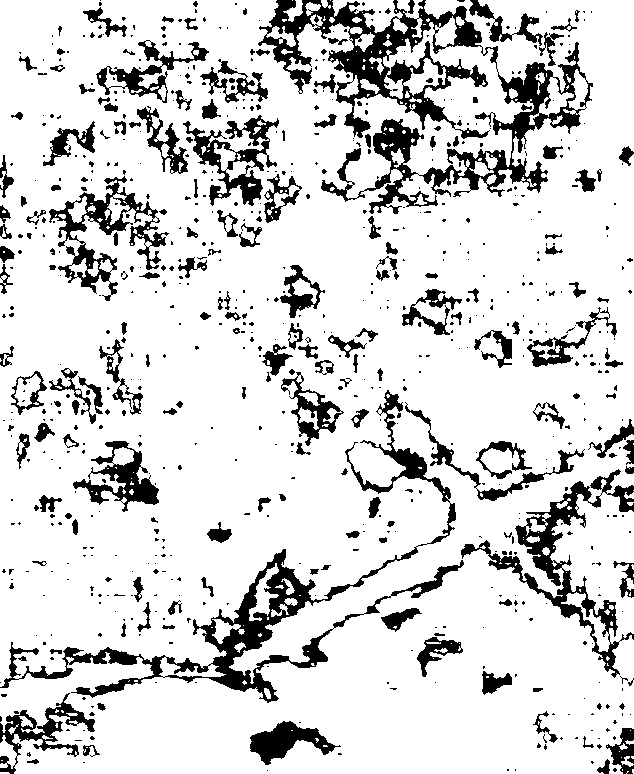}}};

\node (4) [right of=2, xshift=3.5cm] {\fbox{\includegraphics[width=25pt]{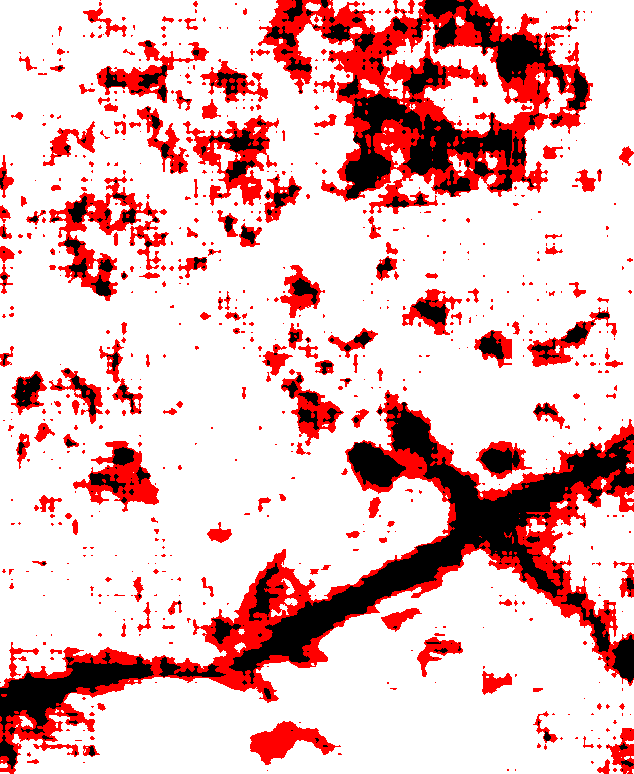}}};

\node (ancillaryAbove3a) [below of=3a, yshift=1.70cm] {CD output};
\node (ancillaryBelow3b) [below of=3b, yshift=0.28cm] {Confidence map};
\node (ancillaryBeside4) [rotate=90,right of=4, yshift=-0.7cm, xshift = -1cm] {Reliable CD output};

\draw [arrow] (1) -- (2);
\draw [arrow] (2) -- (3a);
\draw [arrow] (2) -- (3b);
\draw [arrow] (3a) -- (4);
\draw [arrow] (3b) -- (4);

\end{tikzpicture}
 \caption{The key idea: proposed method processes pre and post-change images to obtain a CD output (black - changed, white - unchanged). Proposed method also generates a confidence map (white - confident, black - not confident). Fusion of them produces a reliable CD output (black - confidently changed, white - confidently unchanged, red-  not confident).}
 \label{figureConceptOfUnsupCDUncertainty}
\end{figure*}

Despite the above-mentioned benefits provided by the deep transfer learning based CD methods, they  show significant variation in performance depending on the target scenes or datasets. Furthermore, their performance is dependent on affinity between the target scene and the  dataset on which the pre-trained network is trained \cite{saha2020building}. They also show some dependence on the architecture of the pre-trained network. In a real-life application, it might not be always possible to obtain a pre-trained network that has characteristics suitable for the target area. Relying on the sub-optimal results obtained using deep transfer learning based methods may be risky for safety-critical applications, e.g., disaster response. In such cases, it might be more desirable to have confident output over a subset of pixels, instead of having uncertain output over the entire target scene. 
\par
The aforementioned desire to achieve more certain results is not confined solely to change detection; it extends to other problems of Earth observation, such as image classification or semantic segmentation. Remote sensing community has shown increased interest towards uncertainty quantification in the last few years \cite{gawlikowski2022advanced}. One significant factor contributing to the anomalous behavior of deep learning models in Earth observation tasks is the presence of out-of-distribution (OOD) data during the testing phase. OOD data refers to data points or instances that are substantially different from the in-domain data on which the model was trained.  Gawlikowski \textit{et al.} \cite{gawlikowski2022advanced} proposed a method for supervised OOD detection for supervised image classification. Here, supervised OOD detection denotes that there is an additional labeled pool of OOD data in addition to the labeled data from the in-domain dataset. However, the challenges addressed in our work, which is deep transfer learning-based change detection, is different from traditional OOD detection. Since our task does not involve any training process, there is no explicit notion of distributional uncertainty between training and test data. Our objective is not only to distinguish changed pixels from unchanged ones but also to determine which pixels' output can be considered as more reliable. This combines two tasks: the primary task of identifying changes and the auxiliary task of identifying confident predictions, all in an unsupervised manner. Solving this problem is important for enhancing the overall performance and interpretability of change detection models by empowering the model to not only recognize changes but also communicate its level of confidence for those predictions, as illustrated in Figure \ref{figureConceptOfUnsupCDUncertainty}.
\par
Randomized smoothing is a transformation of a classifier, that is a popular technique in the context of adversarial robustness \cite{cohen2019certified}. It allows us to determine the most likely output for an input perturbed by noise. Taking inspiration from this paradigm, we postulate that deep transfer learning based CD can employ two pre-trained networks simultaneously, one tasked to identify the changed pixels and the other tasked to obtain the confidence of output by ingesting ensembles of input perturbed by noise. \par
The key contributions of this paper are as follows:
\begin{enumerate}
\item We propose an unsupervised method for determining the pixels for which the outcome of the deep transfer learning based CD is more confident. To the best of our knowledge, no existing deep unsupervised CD method can provide such confidence score. 
\item We introduce the key concept of randomized smoothing to deep unsupervised change detection.
\item We demonstrate the proposed approach for three different types of sensors, multi-spectral, hyperspectral, and Synthetic Aperture Radar (SAR) sensors. This shows the versatility of the proposed approach for different input conditions.
\end{enumerate}

\section{Related Works}
\label{sectionRelatedWorks}
Considering relevance to our work, we briefly discuss change detection (in general), transfer learning based deep unsupervised, and pseudo-labeling based deep unsupervised change detection methods. Additionally, we discuss uncertainty quantification in deep learning (DL)-based change detection and randomized smoothing.

\subsection{Change detection}
Both supervised and unsupervised methods have been proposed in the literature for change detection. A significant number of supervised methods are variants of Siamese architectures \cite{daudt2018fully} where twin subnetworks share the same architecture and parameters and ingest one input image each. After feature extraction, the outputs from the twin networks are concatenated or combined in some way to form a single representation for comparison. Instead of using Siamese architectures, some supervised methods combine the inputs and process them using a fully convolutional network \cite{peng2020optical}.  
Similar to semantic segmentation in computer vision, many variations of architecture and loss functions have been used in supervised change detection \cite{li2022transunetcd,li2021combined}. Despite their good performance when suitable training data is available, supervised methods often do not generalize well to geographic and sensory differences, which are common in Earth observation data. Furthermore, unsupervised change detection methods may perform better than the supervised ones when the annotated data is scarce \cite{saha2021trusting}. Owing to this, unsupervised  methods are still popular in  change detection. Two variants of unsupervised deep learning based methods are popular, one that somehow uses a model trained previously for some other task (transfer learning based CD). The other variant somehow pseudolabels some data and then uses those pseudolabels to train a model in supervised manner (pseudo-label based CD).
\par
Among the deep transfer learning based methods, a popular method is DCVA \cite{saha2019unsupervised} that extracts bi-temporal features using pre-trained network and stacks them into a deep change hypervector that is further processed for binary and multiple change detection. It is mention worthy that DCVA itself is similar in principle to the Siamese methods as both process pre-change and post-change inputs through weight-sharing twin networks, however different in several aspects including that DCVA does not involve any training. Use of both sensor-specific pre-trained model \cite{saha2020building} and ImageNet-based pre-trained model \cite{pomente2018sentinel} have been attempted in the literature. Several works have exploited the potential of replacing the pre-trained network with networks derived from the data itself, e.g., using self-supervised learning \cite{saha2020unsupervised} or using untrained models \cite{saha2021change}. 
\par
Among the pseudo-label based methods, Li \textit{et al.} \cite{li2019deep} uses spatial fuzzy clustering to generate pseudo labels that are then used to train a  change detection model in supervised manner. Gao \textit{et al.} employ a similar approach for sea ice change detection \cite{gao2019sea}. 
\par
In addition to supervised and unsupervised approaches, other variants also exist, e.g., semi-supervised change detection methods that use both unlabeled data and few labeled samples. Several methods have been proposed in this context, e.g., using graph neural network \cite{saha2020semisupervised} and multi-task learning \cite{shu2022mtcnet}. Pseudo-label based CD is combined with semisupervised techniques in \cite{kondmann2023semisiroc}.

\subsection{Notion of uncertainty in change detection}
The notion of uncertainty in change detection is often used in the works that employ pre-classification for change detection. As an example, Li \textit{et al.} \cite{li2019deep} use a pre-classification scheme to obtain initial clusters of changed and unchanged pixel with a degree of membership (similar to confidence) assigned to each pixel. Similarly, Zhan \textit{et al.} \cite{zhan2020unsupervised} use initial classification of changing superpixels that are categorized by uncertainty and used to train a supervised classifier. A systematic study on trustworthiness of supervised vs. unsupervised change detection models is conducted in \cite{saha2021trusting}.  The idea that a form of ensembles could be used to produce uncertainty score for unsupervised hyperspectral CD is discussed in \cite{saha2022fusing}.

\subsection{Randomized smoothing} 
Deep neural network based models are often found to be susceptible to minor, however adversarial perturbations. Many certification techniques, such as randomized smoothing, have been proposed in this context as a certified defense against such perturbations. Unlike deterministic certification methods, randomized smoothing \cite{cohen2019certified}
provides robustness certification with probabilistic guarantees. Though there may be variation in the functioning of randomized smoothing, in following we describe how it produces certified $\ell_2$ robustness.
For a given classification model, $f$, that maps inputs in $\mathbb{R}^d$ to $\mathcal{Y}$ classes, the randomized smoothing framework transforms the $f$ into a smoothed classifier $h$.
For a given input $x \in \mathbb{R}^d$, the smoothed classifier $h$ returns the most probable class under isotropic Gaussian noises. Though our work is not related to adversarial robustness, we model the uncertainty quantification for our problem statement by taking cues from randomized smoothing.

\par
Proposed work directly extends the DCVA \cite{saha2019unsupervised}, a transfer learning based change detection method, incorporating uncertainty notions into it. However, any other method could also be extended following similar principle with minimal modifications. To incorporate uncertainty notions into DCVA, proposed method uses concepts similar to randomized smoothing \cite{cohen2019certified}. 
Proposed method is also somewhat similar to deep ensembles \cite{lakshminarayanan2017simple}, however different in its use of primary and secondary model. Furthermore, the confident pixels chosen by the proposed method can be used as pseudo-label to train supervised architectures similar to pseudo-label based change detection methods \cite{li2019deep,zhan2020unsupervised}.

\section{Proposed Method}

\label{sectionProposedMethod}
Let us assume that we have a pair of co-registered bi-temporal images $X_1$ and $X_2$. No additional change detection dataset or any training label is available with us. However, we have access to suitable image feature extractor models $\mathcal{F}_1$ and $\mathcal{F}_2$, which might have been originally trained for some other tasks, e.g., image classification or semantic segmentation on different datasets. Our task is to mark each pixel in the given scene as either changed or unchanged and simultaneously classify each pixel as confident or not confident. While the bi-temporal features extracted using $\mathcal{F}_1$ are compared using DCVA \cite{saha2019unsupervised} to identify changed/unchanged pixels, the other model $\mathcal{F}_2$ is used to compute confidence using a strategy inspired from randomized smoothing. The way in which $\mathcal{F}_1$ and $\mathcal{F}_2$ are employed designates them as the primary and secondary models in the proposed method.

\tikzset{
    between/.style args={#1 and #2}{
         at = ($(#1)!0.5!(#2)$)
    }
}

\begin{figure*}
 \centering
 \scalebox{0.85}
 {
 \begin{tikzpicture}
\node (1a) [draw, io, align=center] {Pre-change and post-change images\\ ($X_1$ and $X_2$)};

\node (2b) [draw, align=center, right of=1a,xshift=4.5cm,yshift=-1.5cm] {Noisy\\ $X_1^{1}$ and $X_2^{1}$};
\node (2c) [draw, align=center, right of=1a,xshift=11cm,yshift=-1.5cm] {Noisy\\ $X_1^{K}$ and $X_2^{K}$};

\node (3a) [rounded corners=3pt,draw, align=center, below of=1a, yshift=-1.75cm] {DCVA with $\mathcal{F}_1$ for binary CD \\ \textcolor{blue}{Extract features from $X_1$ and $X_2$}
\\ \textcolor{blue}{Compare features}
\\ \textcolor{blue}{Obtain deep change mag. for each pixel}
\\\textcolor{blue}{Assign each pixel to changed/unchanged}};
\node (3b) [rounded corners=3pt,draw, align=center, below of=2b, yshift=-0.85cm] {DCVA with $\mathcal{F}_2$};
\node (3c) [rounded corners=3pt,draw, align=center, below of=2c, yshift=-0.85cm] {DCVA with $\mathcal{F}_2$};

\node (4a) [draw, io, align=center, below of=3a, yshift=-1.45cm] {Primary \\CD output};

\node (4b) [rounded corners=3pt,draw, align=center, below of=3b, yshift=-0.85cm,xshift=3.2cm] {Count $K'$ for each pixel\\ from all noisy inputs};
\node (5) [rounded corners=3pt,draw, align=center, below of=4b, yshift=-1.35cm] {Compare whether primary CD output and \\ $K'$ computed using $\mathcal{F}_2$ reasonably agree
\\
\textcolor{blue}{Confident: $K' \geq K_{\tau}*K$ and primary output is changed}
\\
\textcolor{blue}{or}
\\
\textcolor{blue}{Confident: $(K-K') \geq K_{\tau}*K$ and primary output is unchanged}
\\
\textcolor{blue}{or}
\\
\textcolor{blue}{Otherwise not confident}};
\node (6) [draw, io, align=center, below of=5, yshift=-1.45cm] {Confidence map};

\draw [arrow] (1a) -- (3a);
\draw [arrow] (2b) -- (3b);
\draw [arrow] (2c) -- (3c);
\draw [arrow] (3a) -- (4a);
\draw [arrow] (3b) -- (4b);
\draw [arrow] (3c) -- (4b);
\draw [arrow] (4b) -- (5);
\draw [arrow] (5) -- (6);

\node (pointTopOf2b) [shape=coordinate] [below of=2b, yshift=2.43cm] {};
\draw[arrow](1a) -- (pointTopOf2b) -- (2b);

\node (pointTopOf2c) [shape=coordinate] [below of=2c, yshift=2.4cm] {};
\draw[arrow](1a) -- (pointTopOf2c) -- (2c);

\node (dotsBetween2bAnd2c)  [right of=2b, xshift=2.1cm] {.  .  .};
\node (dotsBetween3bAnd3c)  [right of=3b, xshift=2.1cm] {.  .  .};

\node (pointLeftOf5) [shape=coordinate] [left of=5, xshift=-7.7cm] {};
\draw[arrow](4a) -- (pointLeftOf5) -- (5);

\node (ancillaryRightOf1a) [right of=1a,xshift=3.3cm,yshift=0.22cm] {Add noise};

\end{tikzpicture}
}
 \caption{Proposed mechanism for computing confidence map given $X_1$, $X_2$, $\mathcal{F}_1$, $\mathcal{F}_2$}
 \label{figureDCVAWithConfidenceOutputFlowchart}
\end{figure*}
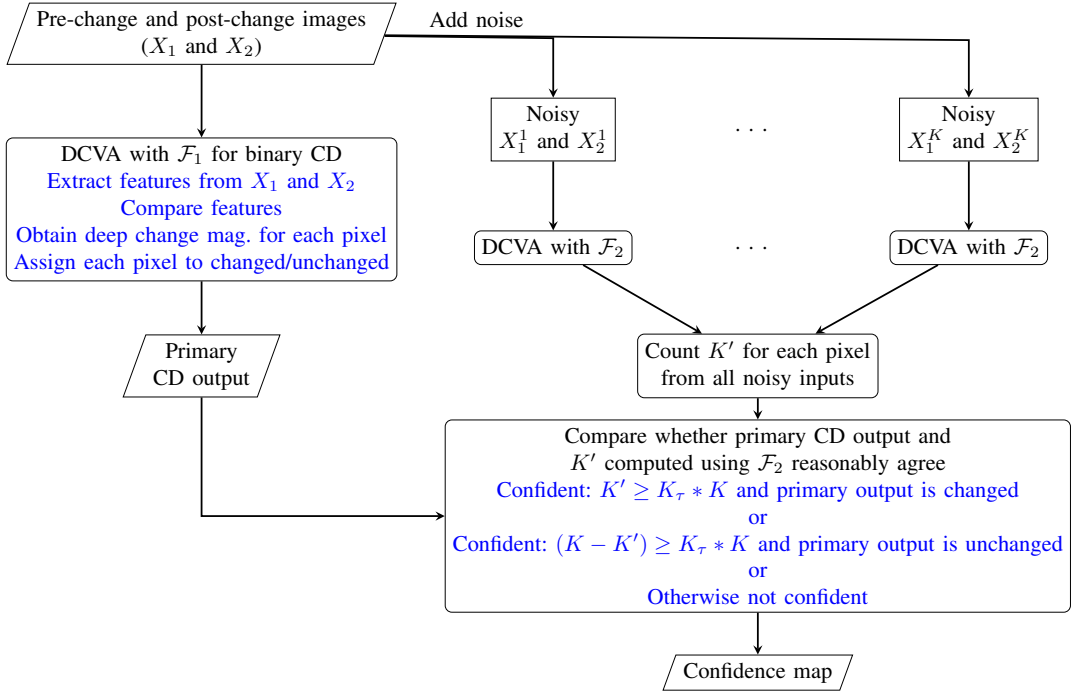

\subsection{Change detection}
\label{subsectionChangeDetection}
The primary model $\mathcal{F}_1$ is applied separately to $X_{1}$ and $X_{2}$ to extract a unique set of deep features for each pixel in the scene, as described in \cite{saha2019unsupervised}. Using the same model for both images ensures that similar inputs (pixels) are mapped to comparable representations in the feature space, while dissimilar pixels are mapped to distinct feature representations. This occurs because they undergo processing through the same set of functions.
Deep features are then extracted from a set of layers of the network to create a pixel-wise deep change hypervector denoted as $G$ \cite{saha2019unsupervised}. This hypervector is derived as the differences in deep features between $X_{1}$ and $X_{2}$. The components of $G$, denoted as $g^{d}$ $(d=1,...,D)$, tend towards zero for unchanged pixels ($\omega_{nc}$), while they tend towards larger values (either positive or negative) for the changed pixels ($\Omega_c$).
To distinguish between $\Omega_c$ and $\omega_{nc}$, we calculate the deep change magnitude denoted by $\rho$ for each pixel, defined as the Euclidean norm of $G$.
The $\rho$ transforms the $D$-dimensional $G$ into a single $1$-dimensional index, while still preserving the key characteristics of the changes. Pixels that remain unchanged typically exhibit lower values of $\rho$, compared to pixels that have undergone a change. 
This inherent property forms the basis for distinguishing between two sets of pixels: $\Omega_c$ and $\omega_{nc}$. To make this discrimination, a threshold, denoted as $\tau$, is employed. This threshold is computed using a thresholding mechanism applied across the entire scene, as outlined by Otsu's method \cite{otsu1979threshold}. While the specific details of the thresholding method are not the central focus of the current study, it is important to acknowledge that different thresholding mechanisms exist. Nevertheless, it has been demonstrated in \cite{saha2021change} that the results obtained using various thresholding mechanisms show minimal variation.  Any pixel with a $\rho$ greater than $\tau$ is assigned to the set of changed pixels ($\Omega_c$), and conversely, pixels with a $\rho$ less than or equal to $\tau$ are assigned to the set of unchanged pixels ($\omega_{nc}$).

\subsection{Confidence assignment}
\label{subsectionConfidenceAssignment}
The secondary model $\mathcal{F}_2$ is used for  confidence computation.  Similar to principles governing randomized smoothing, we introduce variation in our input data, creating multiple perturbed or noisy renditions of $X_1$ and $X_2$. More specifically, Gaussian noise is added similar to \cite{cohen2019certified}. Let us denote one such noisy instance as $X_1^{k}$ and $X_2^{k}$. Processing them using $\mathcal{F}_2$ following same steps as in Section
\ref{subsectionChangeDetection}, we obtain deep change magnitude $\rho$ for each pixel of these perturbed instances, which is thresholded to assign each pixel as changed or unchanged. It's important to note that this entire process is iteratively performed $K$ (number of smoothing iterations) times, each iteration using different instances of perturbed input data. In the iterative process, a particular pixel is identified as changed in a given iteration, let's denote this count as  $K'$, where 
$K'\leq K$. This signifies that out of the 
$K$ times the analysis is performed with different perturbed instances of input images, the pixel in question is detected as changed as $K'$ times. Since different perturbed versions of $X_1$ and $X_2$ are expected to represent the same information as $X_1$ and $X_2$, we expect their output to be same as  change detection output obtained from $X_1$ and $X_2$ using $\mathcal{F}_1$. Ideally, for a pixel to be confidently considered as changed/unchanged, it should be predicted as changed/unchanged both by model $\mathcal{F}_1$ and in $K$ runs using model $\mathcal{F}_2$. More specifically, a pixel predicted as changed by $\mathcal{F}_1$ and obtained value of $K'$ equal to $K$ using $\mathcal{F}_2$, is a confident changed pixel. On the other hand, a pixel predicted as unchanged by $\mathcal{F}_1$ and obtained value of $K'$ equal to $0$ from $\mathcal{F}_2$, is a confident unchanged pixel.
\par
However, in practice, one can expect some variation among results produced in different runs of $\mathcal{F}_2$. So, we introduce an additional hyperparameter, confidence threshold $K_\tau \in (0,1]$ that is an indicator of expectation of consensus among the results obtained from $\mathcal{F}_2$. For a given pixel, if the outputs obtained from $\mathcal{F}_2$ with different noisy input instances agree at least $K_\tau \times K$ times with the output obtained from  $\mathcal{F}_1$, we categorize the pixel's output as confident; otherwise, it is deemed not confident.  Even if the outputs obtained from $\mathcal{F}_2$ exhibit the required level of agreement among themselves, the pixel is not considered confident if this output does not align with the output obtained from $\mathcal{F}_1$.  The alignment or divergence between the outputs obtained using two different models provide insights into the confidence level of the prediction. Another way of looking at this is that the
outputs obtained from secondary model $\mathcal{F}_2$ are merely validating the output obtained from the primary model $\mathcal{F}_1$, similar to the certification process in randomized smoothing.
It's important to emphasize that although our confidence detection approach shares conceptual similarity with randomized smoothing, it does not incorporate any training phase, unlike randomized smoothing \cite{cohen2019certified}.
\par
Figure \ref{figureDCVAWithConfidenceOutputFlowchart} visually summarizes the steps of the proposed confidence map computation.

\subsection{Hyperparameters w.r.t. confidence assignment}
\label{subsectionHyperparameters}
The confidence detection strategy partly depends on a set of hyperparameters. It is well known that hyperparameters are important for functioning of any deep learning model. However, this is more challenging in our case as we do not have any training/validation data and any training/validation phase. Going with the usual assumption of unsupervised change detection, one cannot merely use a validation set to find the best values for such parameters. Furthermore, one set of parameter that works for one dataset may not work for another dataset. In the following we discuss some hyperparameters and some heuristics that can be used to fix their values:
\begin{enumerate}
    \item \textbf{Noise level:} The noise level refers to the magnitude or intensity of the added random noise. The noise level is an essential parameter in randomized smoothing and needs to be selected and fine-tuned based on the nature of the data. Similar to the randomized smoothing, noise level also plays an important role in our case. Cohen \textit{et al.} \cite{cohen2019certified} states that noise level can be higher for higher resolution images, as they can tolerate higher level of noise before class-distinguishing features get destroyed. In our case, we draw noise from Gaussian distribution and use same reasoning as \cite{cohen2019certified} to decide the noise level.

     \item \textbf{Choice of $\mathcal{F}_1$ and $\mathcal{F}_2$}: It is well known that any deep learning system depend on the chosen model architecture. While this is true for our case as well, the proposed method can work for any choice of $\mathcal{F}_1$ and $\mathcal{F}_2$. Given that $\mathcal{F}_1$ is tasked to perform the primary task, i.e., change detection and is applied only once on the bi-temporal images, it is reasonable to choose a deeper architecture as $\mathcal{F}_1$, compared to the architecture for $\mathcal{F}_2$ that needs to be applied repeatedly on the target images. 
     Ideally, one could use the same model for both $\mathcal{F}_1$ and $\mathcal{F}_2$; however, doing so would diminish the variability that the proposed method leverages to its advantage.
    
    \item \textbf{Value of $K$}: Each output of $\mathcal{F}_2$ essentially represents a different realization of random noise added to the actual inputs $X_1$ and $X_2$. The outputs from $K$ runs are computed, and these outputs are subsequently compared with the output from $\mathcal{F}_1$ to compute confidence. Therefore, the choice of the value for $K$ can influence the characteristics of the proposed model. While an exceedingly large value of $K$ would be ideal in theory, it comes at the cost of increased computational demands. To strike a balance between these two factors, it is necessary to select a moderate value for $K$.

    \item \textbf{Confidence threshold ($K_\tau$)}: The choice of the confidence threshold can significantly impact the results. A higher confidence threshold implies that only pixels with very high confidence levels are designated as confident. This approach results in a more conservative selection, yielding a smaller number of confident pixels. While this selection may be characterized by a high degree of reliability, it might also miss out on some confident pixels that are slightly below the very high threshold. Conversely, opting for a lower confidence threshold results in a broader acceptance of pixels as confident. This leads to a larger set of confident pixels, but selected pixels may have lower confidence values, making them less reliable in comparison to those selected with a higher threshold. A higher threshold prioritizes reliability but may overlook some confident pixels, while a lower threshold captures a larger number of confident pixels but with a potential trade-off in reliability. We argue that choice of $K_\tau$ is also dependent on the sensor under consideration. For sensors that capture comparatively noisy images, e.g., SAR, $K_\tau$ value can be kept lower in comparison to its value for optical sensors.
\end{enumerate}

\section{Results}
\label{sectionExperimentalResult}
\subsection{Datasets}
\label{sectionDataset}
We conduct experiments on following datasets:
\begin{enumerate}
\item Onera Satellite Change Detection (OSCD) dataset \cite{daudt2018urban} is popular in the context of unsupervised change detection evaluation \cite{saha2021trusting}. We use the OSCD test set that contains 10 bi-temporal optical multispectral (Sentinel-2) scenes. All scenes are acquired over different locations in the Earth. This dataset is generally considered as challenging and contains complex urban (and surrounding) changes, including buiding changes, road work, flooding, devegetation, and deforestation.
\item The (ascending order) SAR dataset proposed in \cite{saha2022fusing} is a SAR dataset using the same locations as in the above-mentioned optical OSCD dataset. Thus, the test set contains 10 bi-temporal SAR (Sentinel-1) scenes. We will simply refer this dataset as ``OSCD SAR". Previous works \cite{saha2022supervised} show that change detection performance is significantly lower in OSCD SAR compared to its optical version.
\item The Santa Barbara dataset hyperspectral dataset \cite{lopez2018stacked,lopez2019gpu} is a pair of 224 spectral bands, 984 $\times$ 740 pixel images from 2013  and 2014.
\end{enumerate}
Thus, our experiments cover three different Earth observation sensors that show significantly different characteristics.  It is widely acknowledged that methods suitable for one sensor type may not generalize well to others, especially considering the inherent differences between optical and SAR images. Similarly, due to the rich spectral information present in hyperspectral images, specialized algorithms are often necessary. Hence, our experiments deliberately span a broad range of sensor diversity.

\subsection{Compared methods}
We recall from our discussion in Section \ref{sectionIntroduction} that confident pixel identification is relatively new topic in context of deep transfer learning based change detection. To verify the effectiveness of the proposed method, we compare it to following confident pixel detection mechanisms:
\begin{enumerate}
\item Firstly, we compare to the DCVA without any confident pixel detection.
\item As discussed in Section \ref{subsectionHyperparameters}, proposed approach uses two pre-trained models,  one that is tasked for change detection only ($\mathcal{F}_1$) and the other that assists in the confident pixel selection ($\mathcal{F}_2$). However, both models can even be same, i.e, $\mathcal{F}_2$ can be same as $\mathcal{F}_1$. The main limitation here is that it reduces variability in the uncertainty quantification. To thoroughly investigate this hypothesis, we compare the proposed dual-model system to a baseline scenario where $\mathcal{F}_1$ is used for both change detection and confident pixel selection, essentially treating $\mathcal{F}_1$ as $\mathcal{F}_2$. In this configuration, the model is referred to as the "Unified".
\item  The $\mathcal{F}_2$ need not be deep learning model, it can also be traditional change detection model. Thus, we created a baseline using a similar strategy to the proposed method, however instead of $\mathcal{F}_2$ based secondary change detection, using a traditional method, e.g., robust change vector analysis (RCVA) \cite{thonfeld2016robust}.  Henceforth, we refer to this mechanism as simply ``Conf-RCVA".
\item Though originally DCVA and related works do not explicitly explore the idea of uncertainty detection, the concept of confident pixel detection is somewhat embedded within the framework of DCVA \cite{saha2019unsupervised}, through the deep change magnitude parameter, denoted as $\rho$. We established a baseline by exclusively using the conventional DCVA model and employing the deep change magnitude ($\rho$) as a criterion for identifying confident pixels. Changed pixels with high $\rho$ value and unchanged pixels with low $\rho$ value are selected as confident pixels. To implement this, for each pixel in the analyzed scene we derive $\rho'$ as absolute value of difference between its $\rho$ and $\tau$ (threshold between changed pixels and unchanged pixels). Following this, we simply compute a threshold \cite{otsu1979threshold} on $\rho'$ values in the analyzed scene and pixels with $\rho'$ values higher than this threshold are considered confident. Henceforth, we refer to this mechanism as simply ``deep magnitude".
\end{enumerate}

\subsection{Comparison indices}
 Results are compared in terms of precision, accuracy computed over changed pixels (sensitivtiy), accuracy computed over unchanged pixels (specificity), F1 score computed over changed pixels (denoted as F1 ch in Tables), F1 macro (denoted as F1 mac. in Tables) , and percentage of pixels retained as confident by the proposed method. All evaluation indices are shown in the range $0-100$, even if some of their traditional definition might be in the range $0-1$. 
 
\subsection{Feature extraction models}
Models $\mathcal{F}_1$ and $\mathcal{F}_2$ are selected based on the specific characteristics of each sensor, e.g., for OSCD we use ResNet-18 \cite{he2016deep} trained on EuroSAT \cite{helber2019eurosat} and ImageNet \cite{deng2009imagenet}. For OSCD SAR, we use ResNet-18 models trained on BigEarthNet (SAR) \cite{sumbul2021bigearthnet} and MSTAR \cite{huang2020classification}. For hyperspectral, we use the models in \cite{saha2021change}. We emphasize that although the selection of $\mathcal{F}_1$ and $\mathcal{F}_2$ can influence the output of change detection, this work's primary findings are not overly dependent on their specific selection. This work demonstrates that the collaborative use of two feature extractor models, denoted as $\mathcal{F}_1$ and $\mathcal{F}_2$, according to the proposed methodology, allows us to design a confident pixel selection mechanism and improves the reliability of the results compared to the exclusive use of DCVA with $\mathcal{F}_1$. This underlying principle holds true regardless of the exact choice of $\mathcal{F}_1$ and $\mathcal{F}_2$.

 \subsection{Experiments}
\textbf{OSCD:} 
Table \ref{tableOSCD} provides a detailed overview of the experimental results for the OSCD dataset, with a noise standard deviation level set at $0.1$ for the methods that incorporate noise. The proposed method exhibits a significant enhancement in various evaluation metrics compared to DCVA when no confident pixel selection is applied. For instance, when the proposed method employs a single iteration and a confidence threshold of one, the F1 macro score improves by more than 5 points. This improvement, however, comes at the cost of approximately $22.39\%$ pixel loss. Increasing the number of iterations to 10 further enhances the F1 macro score by $3.29$ point, albeit at an additional pixel loss of around 9$\%$. In comparison to the variant of the proposed method with shared model  ("unified" approach), proposed method consistently obtains superior performance. This shows that indeed having some variation between $\mathcal{F}_1$ and $\mathcal{F}_2$ is helpful. Similarly, the proposed method outperforms Conf-RCVA. We observe that this gap between the proposed method and compared methods increase with increasing number of smoothing iterations. This observation holds true also for deep magnitude based baseline, which is also outperformed by the proposed with smoothing iteration set as 10.
\par
\begin{figure}
 \centering
\begin{tikzpicture}
	\begin{axis}[
		xlabel= Noise,
		ylabel=F1 (macro) and pixel $\%$]
	\addplot[color=red,mark=x] coordinates {
		(0.05,68.48)
		(0.10,70.11)
		(0.15,71.42)
		(0.20,72.09)
		(0.25,72.77)
	};
        \addplot[color=blue,mark=x] coordinates {
		(0.05,72.64)
		(0.10,68.63)
		(0.15,65.26)
		(0.20,61.61)
		(0.25,59.03)
	};
	\end{axis}

\node[anchor=south west] at (0.5,1.5) {\begin{tikzpicture}
    \draw[blue,mark=-] plot coordinates {(-8,2.5)};
    \draw[red,mark=-] plot coordinates {(-8,3)};
    \node[anchor=west] at (-7.8,2.5) {Pixel $\%$};
    \node[anchor=west] at (-7.8,3) {F1 (macro)};
    \end{tikzpicture}};

\end{tikzpicture}

 \caption{Performance variation with different levels of noise when confidence threshold is fixed at 1 (OSCD dataset).}
 \label{figureF1VersusNoiseLevel}
\end{figure}
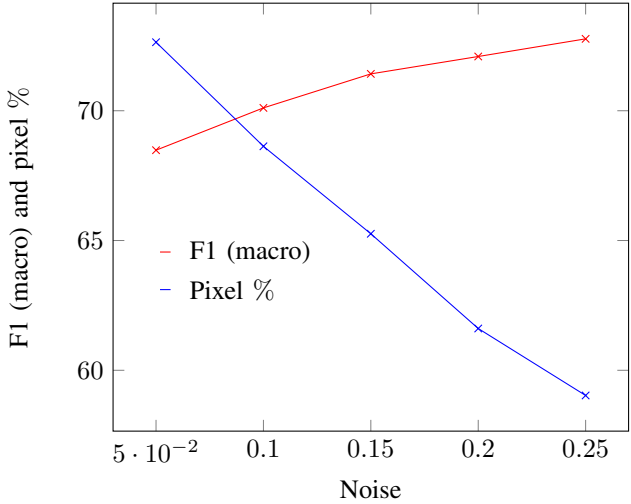

As the number of smoothing iterations increases, the proposed method selects fewer pixels but achieves superior results for the selected pixels. Conversely, when the number of iterations is kept constant (e.g., set to 10), reducing the confidence threshold allows the proposed method to select more pixels, though with a slight reduction in quality. Table \ref{tableOSCD} shows two different values of confidence threshold, 1 and 0.9. Additionally, the variation of F1 macro score with the confidence threshold, while keeping other parameters fixed, is shown in Figure \ref{figureF1VersusSelectionThreshold}. 
Performance experiences a slight boost when introducing more noise during the smoothing process, as illustrated in  Figure \ref{figureF1VersusNoiseLevel}. However, it's worth noting that the number of pixels selected as confident drops significantly when higher noise levels are introduced.

\textbf{OSCD SAR:} 
Table \ref{tableOSCDSar} shows the results on the SAR dataset for two different confidence thresholds with number of smoothing iterations set as 10. The proposed method identifies confident pixels and obtains better accuracy indices over the chosen pixels. However, the performance gain (approx. $5$ point F1 macro) is less prominent in comparison to the performance gain in the optical dataset, which can be attributed to the fact that the change detection in the SAR dataset is relatively challenging \cite{saha2022supervised}. This is also evident from the fact that deep magnitude based baseline performs poorly in comparison to its performance on the optical dataset. Proposed method outperforms two compared methods and obtains similar result to Conf-RCVA.

\textbf{Hyperspectral:}
Table \ref{tableHyperspectral} shows result for a fixed value of smoothing iterations and confidence threshold. Proposed method improves performance over DCVA without confident pixel detection. Though the improvement is less prominent compared to other datasets, partly because DCVA already produces high performance indices. While performance improvement is less prominent, proposed approach also selects higher percentage of pixels for this dataset compared to the other datasets, i.e., the proposed method is confident about its prediction for a large fraction of pixels.  Additionally, proposed method outperforms the compared methods, including unified approach and deep magnitude.

\subsection{Result visualization}
\label{sectionQualitativeVisualization}
\begin{figure*}[!h]
\centering
 \subfigure[]{%
         \fbox{\includegraphics[height=2.5 cm]{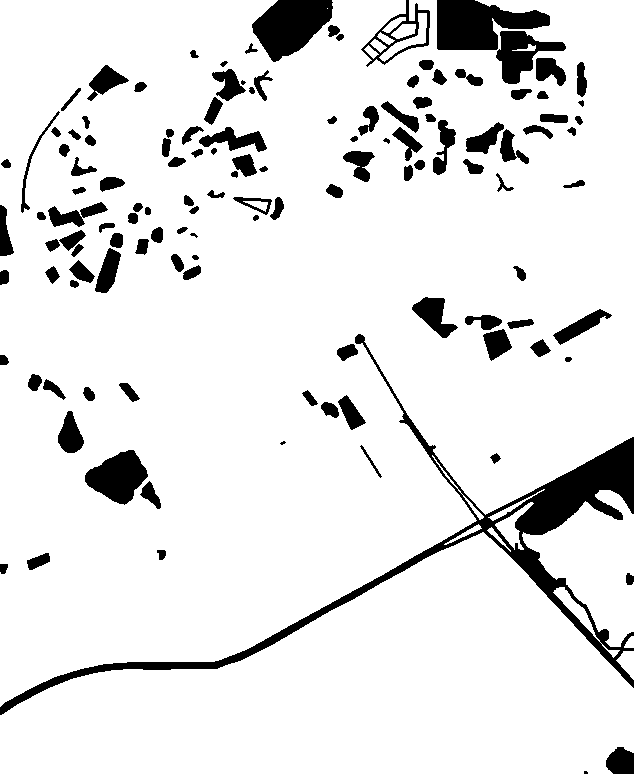}}
            \label{figureSegmentationPolsarScenePauliInput}
           }%
\hspace{0.05 cm}
 \subfigure[]{%
         \fbox{\includegraphics[height=2.5 cm]{figures/figureConfidenceMap/cdMapAfterPixelSelectionDisplaydubai}}
            \label{figureSegmentationPolsarSceneReference}
           }%
\hspace{0.35 cm}
 \subfigure[]{%
         \fbox{\includegraphics[height=2.5 cm]{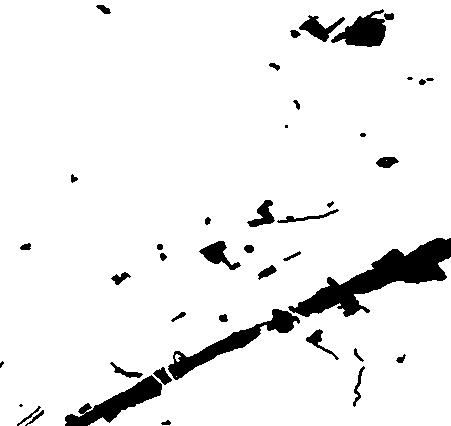}}         
           \label{figureSegmentationPolsarSceneResultProposed}
           }
\hspace{0.05 cm}
 \subfigure[]{%
         \fbox{\includegraphics[height=2.5 cm]{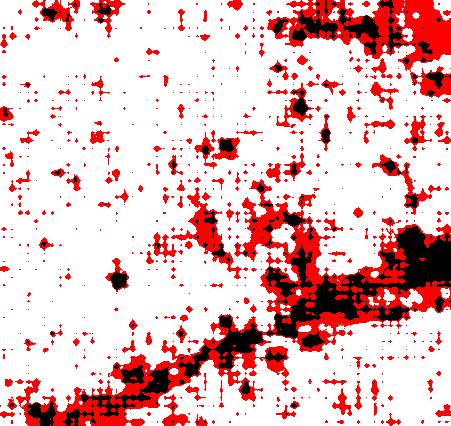}} \label{figureSegmentationPolsarSceneResultProposed}
         }
\hspace{0.35 cm}
 \subfigure[]{%
         \fbox{\includegraphics[height=2.5 cm]{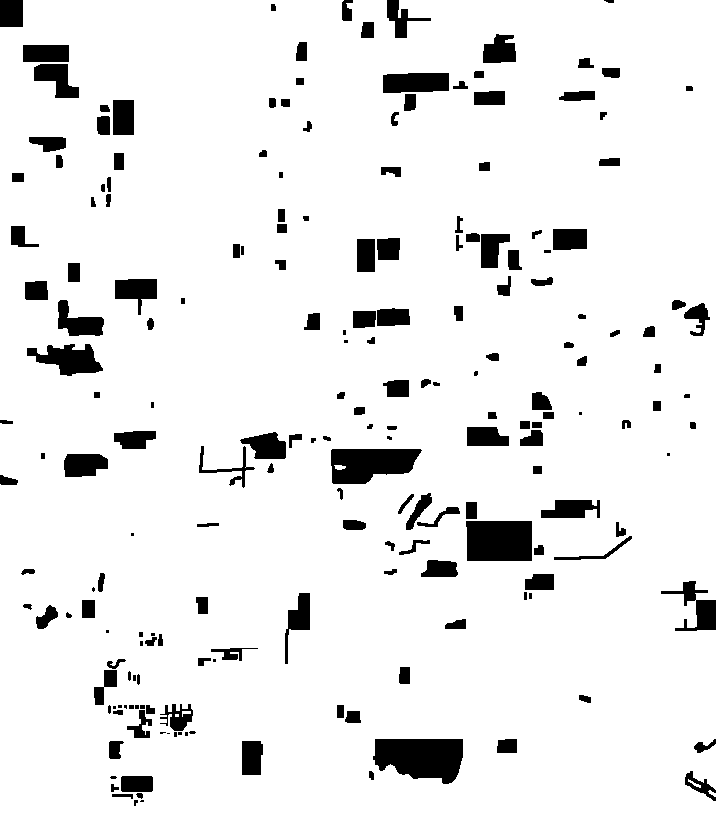}}         
           \label{figureSegmentationPolsarSceneResultProposed}
           }
\hspace{0.05 cm}
 \subfigure[]{%
         \fbox{\includegraphics[height=2.5 cm]{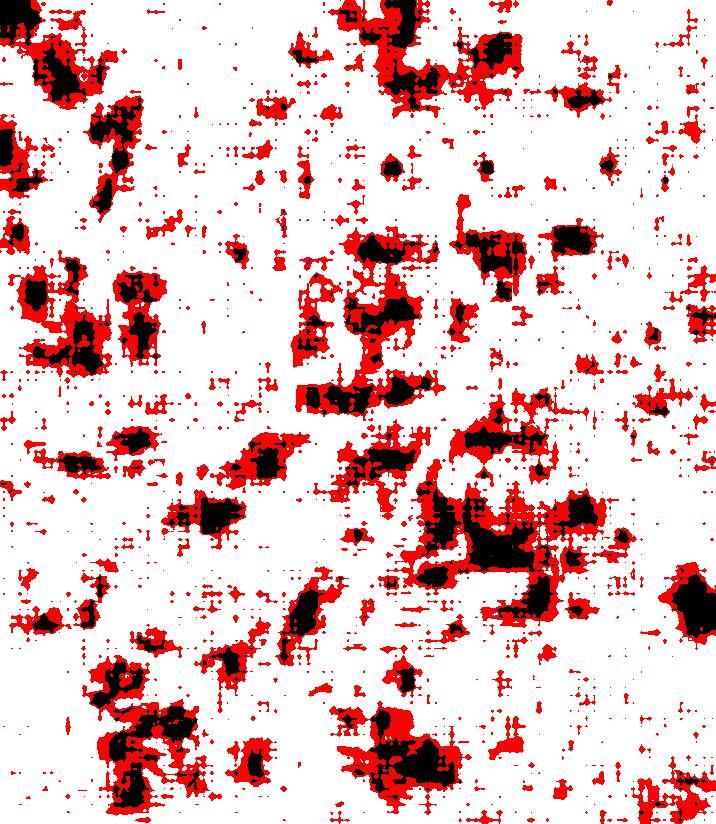}} \label{figureSegmentationPolsarSceneResultProposed}
 }%

\caption{Result visulization on three different scenes (Dubai, Montpellier, Las Vegas, in that order) from OSCD dataset. Subfigures (a),(c),(e) show reference map (black: changed, white: unchanged). Sufigures (b), (d), (f) show corresponding result of the proposed method (black: changed, white: unchanged, red: not confident).}
\label{figureResultVisualizationOscd}
\end{figure*}

Figure \ref{figureResultVisualizationOscd} shows the reference map and result of the proposed method with confident pixel selection for scene of three different scenes (Dubai,Montpellier, Las Vegas) from OSCD dataset. It is evident that the proposed method indeed filters out pixels/objects with wrong prediction, thus improving quantitative result. In addition to many entire isolated objects, proposed method filters out unreliable predictions lying at the edge or boundary of changed/unchanged regions.

\begin{table*}[!h]
\centering
\caption{Quantitative performance of different methods on the OSCD dataset.}
\begin{tabular}{c|l|c|c|c|c|c|c} 
 \hline
\textbf{Parameters} & \textbf{Method} & \textbf{Prec.}& \textbf{Sens.} & \textbf{Spec.} & \textbf{F1 ch.} & \textbf{F1 mac.} & \textbf{Pixel $\%$}    \\ 
\hline
& No conf. est.   & 20.86 & 73.49 & 84.80 & 32.50 & 61.78 & \bf{100} \\
\hline
& Proposed & 26.77 & 72.17 & 90.88 & 39.05 & 66.82 & 77.61  \\
Iterations = 1  & Unified   & 24.89 & 70.94 & 90.14 & 36.85 & 65.50 & 77.92 \\
Conf. thresh. = 1 & Conf-RCVA  & 26.08 & 55.65 & 94.79 & 35.51 & 66.05 & 70.36 \\
\hline
& Proposed & 30.26 & 71.74 & 93.02 & 42.56 & 69.17 & 71.59  \\
Iterations = 5  & Unified   & 28.03 & 70.05 & 92.34 & 40.33 & 67.71 & 71.54  \\
Conf. thresh. = 1 & Conf-RCVA  & 27.54 & 54.00 & 95.47 & 36.47 & 66.71 & 67.46  \\
\hline
& Proposed & \bf{31.84} & 71.61 & 93.67 & \bf{44.08} & \bf{70.11} & 68.63  \\
Iterations = 10  & Unified  & 28.95 & 69.53 & 93.08 & 40.88 & 68.34 & 68.41  \\
Conf. thresh. = 1 & Conf-RCVA  & 28.22 & 53.47 & \bf{95.72} & 36.93 & 67.01 & 66.64  \\
\hline
& Proposed & 30.11 & 71.38 & 92.90 & 42.35 & 69.03 & 73.17  \\
Iterations = 10  & Unified   & 27.37 & 69.69 & 92.20 & 39.29 & 67.30 & 73.26  \\
Conf. thresh. = 0.9 & Conf-RCVA  & 27.21 & 54.14 & 95.36 & 36.21 & 66.55 & 67.88  \\
\hline
& Deep magn.   & 30.15 & \bf{78.82} & 93.07 & 43.61 & 69.81 & 65.40 \\
\end{tabular}
\label{tableOSCD}
\end{table*}

\begin{table*}[!h]
\centering
\caption{Quantitative performance of different methods on the OSCD SAR dataset.}
\begin{tabular}{c|l|c|c|c|c|c|c} 
 \hline
\textbf{Parameters} & \textbf{Method} & \textbf{Prec.}& \textbf{Sens.} & \textbf{Spec.} & \textbf{F1 ch.} & \textbf{F1 mac.} & \textbf{Pixel $\%$}    \\ 
\hline
& No conf. est.   & 12.30 & 48.06 & 81.33 & 19.58 & 53.95 & \bf{100} \\
\hline
& Proposed & 17.73 & 32.99 & 93.22 & \bf{23.06} & \bf{59.04} & 66.84  \\
Iterations = 10  & Unified   & 16.97 & 35.30 & 92.02 & 22.91 & 58.64 & 63.01  \\
Conf. thresh. = 1 & Conf-RCVA  & \bf{20.80} & 21.93 & \bf{96.61} & 21.35 & 59.03 & 66.68   \\
\hline
& Proposed & 17.46 & 33.34 & 92.97 & 22.92 & 58.91 & 67.92  \\
Iterations = 10  & Unified   & 15.90 & 36.79 & 90.85 & 22.20 & 57.97 & 67.91  \\
Conf. thresh. = 0.9 & Conf-RCVA  & 19.82 & 22.95 & 96.21 & 21.27 & 58.90 & 68.86  \\
\hline
& Deep magn.   & 12.30 & \bf{56.41} & 78.42 & 20.20 & 53.48 & 66.77 \\
\end{tabular}
\label{tableOSCDSar}
\end{table*}

\begin{table*}[!h]
\centering
\caption{Quantitative performance of different methods on the hyperspectral dataset.}
\begin{tabular}{l|c|c|c|c|c|c} 
 \hline
\textbf{Method} & \textbf{Prec.}& \textbf{Sens.} & \textbf{Spec.} & \textbf{F1 ch.} & \textbf{F1 mac.} & \textbf{Pixel $\%$}    \\ 
\hline
No conf. est.         & 97.83 & 82.18 & 98.82 & 89.32 & 91.63 & \bf{100} \\
\hline
Proposed              & 98.39 & \bf{84.93} & 99.14 & \bf{91.16} & \bf{93.13} & 91.02  \\
Unified           & 98.58 & 81.48 & 99.27 & 89.21 & 91.69 & 90.59  \\
Conf-RCVA                  & \bf{99.46} & 81.39 & \bf{99.76} & 89.52 & 92.31 & 83.99  \\
Deep magn.        & 98.27 & 83.63 & 99.06 & 90.36 & 92.48 & 91.09
\\
\end{tabular}
\label{tableHyperspectral}
\end{table*}

\begin{figure}
 \centering
\begin{tikzpicture}
	\begin{axis}[
		xlabel= Confidence threshold,
		ylabel=F1 (macro)]
	\addplot[color=red,mark=x] coordinates {
		(1.0,70.11)
		(0.9,69.03)
		(0.8,68.30)
		(0.7,67.71)
		(0.6,67.18)
	};
	\end{axis}
\end{tikzpicture}
 \caption{As less pixels are selected with stronger confidence threshold, accuracy improves (OSCD dataset).}
 \label{figureF1VersusSelectionThreshold}
\end{figure}
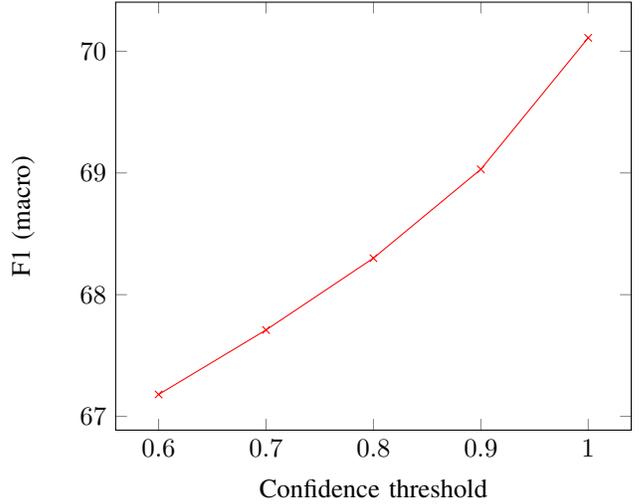

\section{Limitations}

\label{sectionLimitations}
Results of the proposed model may depend on several factors including choice of $\mathcal{F}_1$ and $\mathcal{F}_2$. Like in case of any other change detection model, incorrect result may mislead decision-makers, leading to inappropriate actions. As an example, identifying an area as unchanged when it has actually experienced significant environmental degradation might result in delayed intervention.  However, such risks in our model are reduced in comparison to the majority of existing models that do not account for uncertainty.
\par
One limitation of the proposed method is the relatively modest increase in required computational resources. However, it is essential to recognize that in the majority of applications, change detection tasks are typically performed offline, allowing for a larger allocation of computational resources. In this context, the additional computational expenses incurred by the proposed method are justified, especially when considering the improvement in the reliability and accuracy of the results it provides. Using a Nvidia V100 GPU, our method is capable of processing the each scene of OSCD dataset in a short time frame (50 seconds).

\section{Conclusions}

\label{sectionConclusions}
The concept of quantifying uncertainty in deep models, particularly within the domain of remote sensing, has gained significant attention in recent years. At the same time, change detection stands out as one of the most extensively studied and well-established areas in this field. Our paper resides at the  intersection of these two topics. In our study, we take the innovative approach of harnessing a secondary model designed for confidence estimation. 
\par
Through a series of experiments conducted across different remote sensing sensors, we have demonstrated the effectiveness of our proposed methodology. Our findings show that integrating a confidence estimation model with deep learning techniques enhances the accuracy and reliability of change detection in remote sensing applications. 
\par
The quantitative results of our proposed method typically outperform those of the compared methods, albeit not consistently. It's important to note that our method represents the first example of a deep unsupervised CD method that generates confidence scores. Additionally, the compared methods have been somewhat tailored by us for this task, drawing upon suitable techniques from existing literature. Therefore, it's important not to solely assess the novelty and significance of our work based on quantitative comparisons alone.
\par
 In future, we would like to investigate potential of using confident pixels chosen from the proposed method as pseudo samples to train supervised-like change detection models. The choice of the machine learning-based classifier model to train the model is flexible and depends on the the nature of the dataset. Furthermore, we envision extending the proposed method to tackle the challenges of confidence estimation in semantic (multiple) change detection \cite{zheng2022changemask,saha2019unsupervised}.  This research direction holds promise for improving the accuracy and interpretability of task-specific change detection, opening up new possibilities for a wide range of applications, from urban planning to environmental conservation. Additionally, we aim to explore the potential of offering a probabilistic guarantee for our confidence map.


\ifCLASSOPTIONcaptionsoff
  \newpage
\fi

\bibliographystyle{IEEEtran}
\bibliography{sigproc}

\end{document}